\documentclass[titlepage]{article}
\usepackage[utf8]{inputenc}
\usepackage{amsmath}
\usepackage{hyperref}
\usepackage{xfrac}
\usepackage[dvipsnames]{xcolor}
\usepackage{float}
\usepackage{graphicx}
\usepackage{caption}
\usepackage{subcaption}
\usepackage{array}
\usepackage{xr}
\usepackage{url}
\usepackage{authblk}
\usepackage{cleveref}
\usepackage{multirow}
\usepackage{geometry}
\usepackage{setspace} 
\usepackage[version=4]{mhchem}

\usepackage[numbers]{natbib}

\begin{document}

\title{Designing semiconductor-electrochemical junctions for bioinspired energy transduction}

\author[1]{Jonathon L. Yuly\footnote{email: \href{mailto:name@name.com}{jonathon.yuly@princeton.edu}}}
\affil[1]{Lewis-Sigler Institute for Integrative Genomics, Princeton University, Princeton, NJ, 08540}

\date{\today}

\maketitle

\begin{abstract}
Long ago, life discovered how to efficiently push electrons thermodynamically uphill to lower potential by harnessing energy released by an equal number of electrons moving downhill.  Known as electron bifurcation, this form of energy transduction has never been observed in the absence of natural enzymes. To successfully bifurcate electrons, a system must block short-circuit electron transfers that allow all electrons to flow downhill, while maintaining productive reactions. It is difficult to design systems that catalyze these highly-selective electron flows while minimizing free energy dissipation. Using theories of electron transfer and charge transport, I introduce semiconductor-electrolyte junctions that spontaneously bifurcate electrons analogously to natural enzymes (bifurcating junctions). I simulate a simple but illustrative bifurcating junction with typical material properties, and discuss how more complicated designs could achieve higher performance.
\\
\\

\textbf{Significance Statement}
Electron bifurcation is a process catalyzed by enzymes that can push electrons thermodynamically uphill to higher energy by leveraging the downhill flow of other electrons. This process is difficult to engineer because reactions that result in all electrons flowing downhill (short-circuits) must be prevented. Taking inspiration from natural enzymes, this paper designs tailored semiconductor electrochemical junctions to bifurcate electrons and block short-circuiting analogous to nature's enzymes. These junctions (or future iterations on them) might be used to improve batteries, solar cells, or electrocatalysis.

\end{abstract}

%\begin{multicols}{2}

Living cells efficiently convert energy at the nanoscale using many energy transduction mechanisms. Electron bifurcation is one such strategy \cite{peters2016electron, buckel2018flavin, muller2018electron, buckel2018flavin2, nitschke2012redox}. Electron bifurcating enzymes oxidize a two-electron donor molecule, using the resulting two electrons to reduce separate one-electron acceptor molecules. These electron transfers are coupled, so one electron is pushed thermodynamically uphill to lower potential, using energy released by the other electron. Continual turnover of this cycle produces a steady uphill flow of electrons into a low-potential redox pool and another downhill flow into a high-potential pool. Remarkably, biological electron bifurcation can occur in a near-reversible regime (i.e. with small driving force\cite{fourmond2021reversible}), allowing near-100\% energy transduction efficiency\cite{buckel2018flavin2, yuly2019electron}. 

Because biological electron bifurcation efficiently concentrates the reducing power (redox potential energy) of electrons, there is growing interest in bioinspired electron bifurcation for applications in catalysis\cite{yuly2019electron,arrigoni2020rational, das2020biological, yuly2020universal, huang2025design}. However, bioinspired electron bifurcation has never been demonstrated (i.e. without using natural enzymes). A molecular bifurcating system must be designed with extraordinary finesse, and must accomplish selective multi-electron transfer while simultaneously suppressing short-circuiting electron transfers. If such systems could be designed and widely deployed, they might be used to boost the open-circuit voltage of batteries or solar cells, or drive difficult redox reactions at lower overpotential. Simply, electron bifurcation concentrates reducing power into fewer electrons.

Must electron bifurcation occur at the protein scale? If larger constructs could be designed to bifurcate electrons, then they could be fabricated using ``top-down'' methods with precision. In this paper, inspired by a mechanism that enables electron bifurcation in enzymes\cite{yuly2020universal}, I introduce the design of electron bifurcating semiconductor-electrolyte junctions, or bifurcating junctions. Section 1 briefly reviews the mechanism of electron bifurcation that is proposed in Ref \cite{yuly2020universal}, drawing an analogy with the physics of junctions between electron (n) and hole (p) doped semiconductors. Section 2 introduces bifurcating junctions and outlines how they work. Section 3 describes a detailed simulation of a simple bifurcating junction. Finally, the last section proposes concepts for future designs with higher performance or particular applications.

\section{Biological electron bifurcation}

\begin{figure}[H]\centering
\includegraphics[scale=0.75]{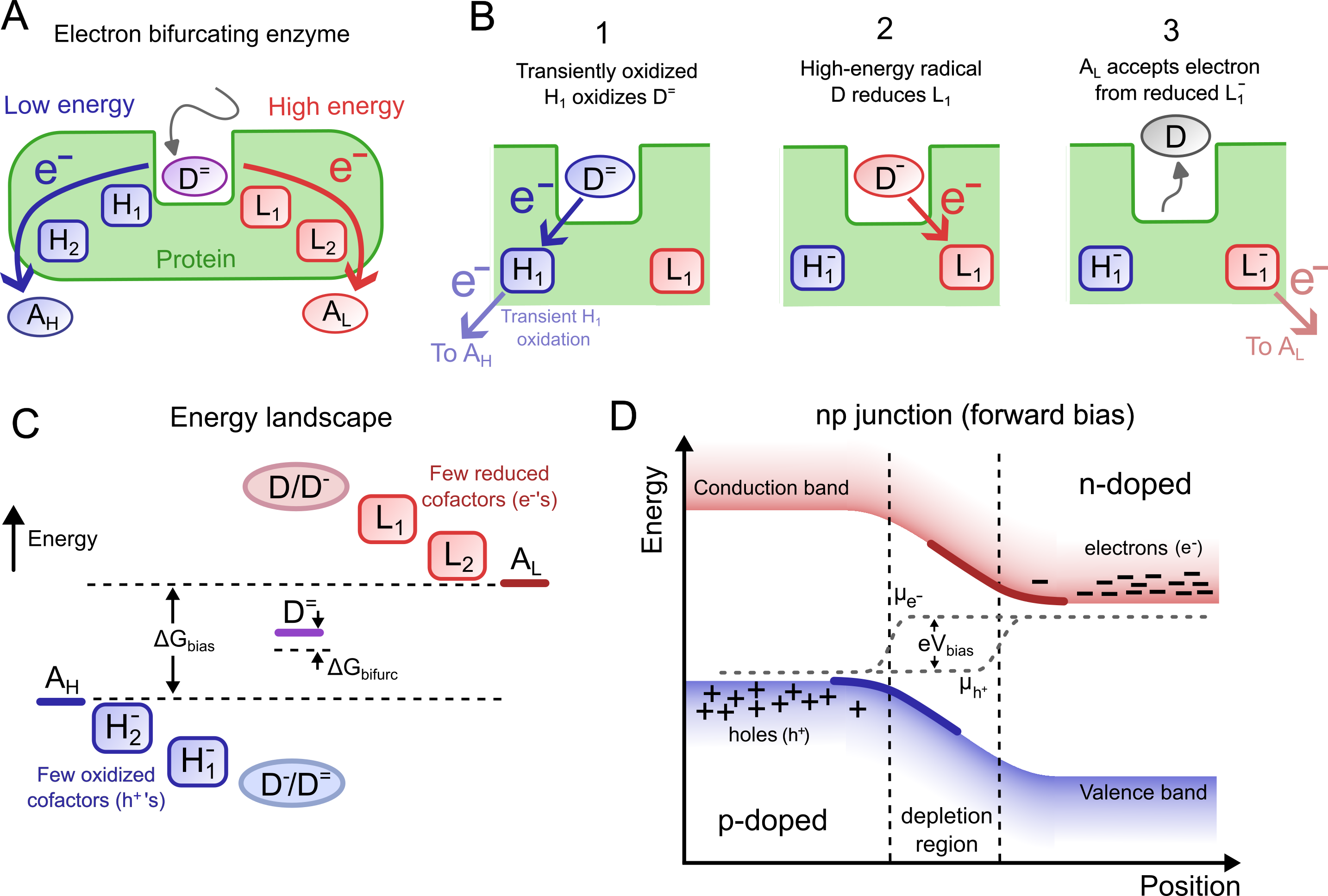}
\caption{\label{figure1} \textbf{Overview of biological electron bifurcation and energy landscape analogy with semiconductor np junctions.} (A) A simple electron bifurcating enzyme oxidizes a two electron donor molecule D, shuttling each electron down a different ``branch'' of cofactors to reduce two final acceptors, one at high-potential ($\ce{A_H}$) and the other at low-potential ($\ce{A_L}$). (B) When operating in the bifurcating (forward) direction, the electron donor $\ce{D^=}$ undergoes a series of redox reactions when bound to the active site (protons are not shown for simplicity). (1) First, the first cofactor in the high-potential branch is transiently oxidized by a thermal (i.e. a ``hole'' moves from the $\ce{A_H}$ pool to $\ce{H_1}$. This allows $\ce{H_1}$ to oxidize $\ce{D^=}$ forming the highly reactive (often radical) species $\ce{D^-}$ that can (2) inject a high-energy electron into the low-potential branch. The second electron does not follow the first electron to the high-potential branch because $\ce{H_1}$ is no longer oxidized ($\ce{H_1}$ is maintained in a reduced $\ce{H^-_1}$ state by the $\ce{A_H}$ pool). Lastly, the high-energy electron is pulled out of the low-potential branch to reduce $\ce{A_L}$. (C) An energy landscape with steep (hundreds of meV) energy landscapes maintains the cofactors in redox states that prevent short-circuit electron transfers. The low-potential branch is maintained in an oxidized state (few ``holes'' to accept short circuit electrons) and the low-potential branch is maintained in an oxidized state (few electrons available to short circuit). The $\ce{A_L}$ and $\ce{A_H}$ redox pools are out of equilibrium (positive $\Delta G_{\text{bias}}$), so electrons from the $\ce{D^=}$ pool flow thermodynamically uphill to the $\ce{A_L}$ pool. (D) There is a strong analogy between the electron energy landscape in (C) and the band bending in the depletion region of np semiconductor junctions, with the same suppression of charge carriers. An applied bias $|e^-|V_{\text{bias}}$ between electron and hole quasi-fermi levels $\mu_{e^-}$ and $\mu_{h^+}$ is analogous to $\Delta G_{bias}$ in the enzyme energy landscape.}
\end{figure}

The function of a generic electron bifurcating enzyme is depicted in Figure 2A. A free two-electron donor molecule ($\ce{D^=}$) diffuses to the site where electron bifurcation occurs. There $\ce{D^=}$ is oxidized, and the resulting electrons quantum mechanically tunnel through two separate paths (called branches) of redox cofactors embedded in the protein matrix ($\ce{L_1}$, $\ce{L_2}$, $\ce{H_1}$, and $\ce{H_2}$). At the terminus of these redox branches, the electrons have access to final one-electron acceptor pools at high and low potentials ($\ce{A_H}$ and $\ce{A_L}$ respectively). Electron bifurcation can be fully reversible, so both directions of electron transfer are relevant between each pair of cofactors in each branch\cite{osyczka2004reversible, osyczka2005fixing, yuly2020universal}. The system in Figure \ref{figure1} is generic for illustration; electron bifurcating enzymes in nature are remarkably diverse\cite{feng2024structures, appel2021functional, watanabe2021three}.

To accomplish robust energy transduction, electron-bifurcating systems must defeat energy-wasting short circuit reactions that uncouple the uphill and downhill electron flows and would cause all electrons to flow downhill\cite{osyczka2004reversible,sarewicz2021catalytic, crofts2013mechanism, yuly2020universal, yuly2021efficient}. It is not enough that the electrons are successfully injected into the high- and low-potential branches by the reactions of Figure \ref{figure1}B, because the kinetics of the electron flows inside the branches must prevent successfully injected electrons from flowing back to the bifurcating site to short circuit\cite{yuly2021energy,osyczka2004reversible, osyczka2005fixing}. 

Recently, it was discovered that high-efficiency electron bifurcation could spontaneously emerge when steep energy gradients exist in both branches (scale of several hundred meV)\cite{yuly2020universal}. This effect is illustrated in Figure 2C. With steep energy gradients, electrons pile up in the high-potential branch, while the low-potential branch remains dominantly oxidized (filled with ``holes''). Thus, electrons are rarely available to short circuit and, even when they are, cofactors in the high-potential branch are rarely in an oxidized state ready to accept short-circuiting electrons. A productive electron bifurcation event is initiated by a sequence of electron transfer reactions at the bifurcating site (Figure 1B). First, the proximal cofactor on the high-potential branch is transiently oxidized, allowing it to in turn perform a one-electron oxidation of $\ce{DH^-}$, yielding the high energy $\ce{D^{\cdot-}}$ species. This species has sufficient energy to reduce the low-potential branch, where the electron flows to the final low-potential acceptor $\ce{A_L}$

As the magnitude of the energy gradients in the branches are increased, the short-circuiting turnover can be made arbitrarily small compared to the productive turnover\cite{yuly2020universal, yuly2021energy}, enabling high-efficiency electron bifurcation. Note that the energy landscape need not be uniformly increasing or decreasing as illustrated in Figure 2C (indeed ``bumps" have been found in the redox energy landscape of some electron bifurcating enzymes\cite{wise2022uncharacteristically}). The constraint is only that the free energy to place hazardous cofactors (those within electron transfer range to short circuit) into short-circuiting redox states must be sufficiently large (hundreds of meV).

There is an analogy between the energy landscape of Figure \ref{figure1}C and the energy band structure found at the interface between n-doped (to introduce electrons) and p-doped (to introduce holes) semiconducting material\cite{yuly2020universal}. At such n-p junctions, a space charge spontaneously appears in the depletion region as both electron and hole carriers are suppressed at the interface between the two doped materials. This space charge generates an electric field across the interface that shifts the energies of the valence and conduction bands continuously throughout the depletion region, as illustrated in Fig 2D. Electrons and holes are both pushed out of the depletion region by the electric field. However, some carriers can occasionally penetrate, with more electrons towards the n-doped side and more holes toward the p-doped side\cite{sze2021physics}.   
These small carrier concentrations (compared with the carrier concentrations deep into the n- and p-doped regions) are analogous to the transient oxidation (reduction) of the $\ce{H_1}$ and $\ce{L_1}$ redox sites in the bifurcating enzyme of Figure 2A. The next section demonstrates how this analogy between energy landscapes can be used to design bifurcating semiconductor-electrolyte junctions.

\section{Electron-bifurcating semiconductor-electrolyte junctions}

\begin{figure}[H]\centering
\includegraphics[scale=0.8]{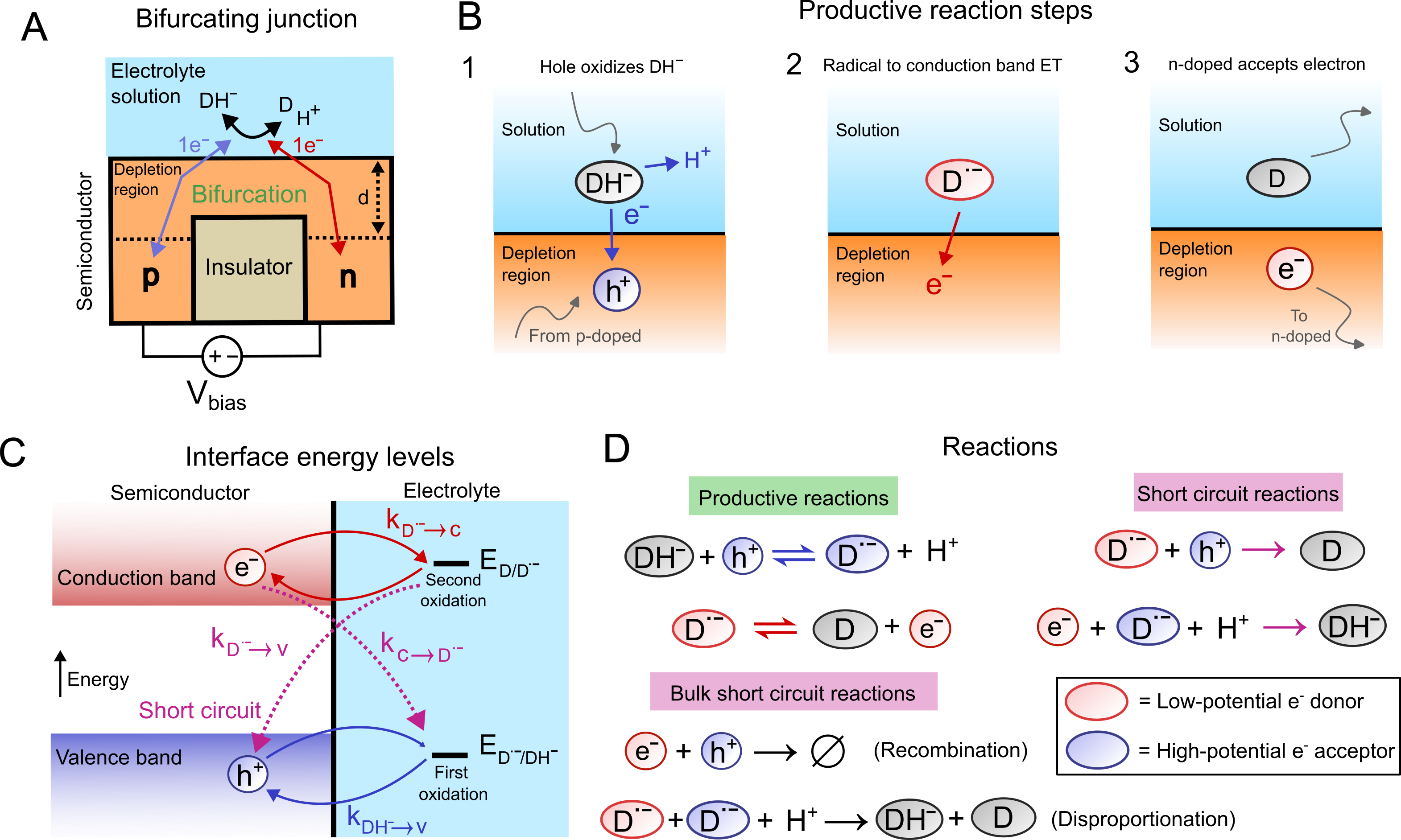}
\caption{\label{figure2} \textbf{Proposed electron-bifurcating semiconductor-electrolyte junction} (A) An electron-bifurcating junction  uses a tailored semiconducting electrode to spontaneously bifurcate electrons from a two-electron species ($\ce{DH^-}$) toward p- and n-doped regions with a voltage $V_{\text{bias}}$ applied across them (if doping is undesired. (B) Interfacial electron transfer reactions that result in spontaneous electron bifurcation across the n-p junction. (a) First, a thermal fluctuation brings a hole ($\ce{h+}$) from the p-doped region to the surface of the depletion region, and performs a one-electron oxidation of $\ce{DH^-}$ to the radial $\ce{D^{\cdot -}}$ species. (2) This radical species has sufficient reducing power to inject a charge into the conduction band. (3) Finally, the electron flows to the n-doped region while oxidized $\ce{D}$ diffuses back to the counter electrode to be refilled with electrons. (C) On the left is shown the energy of the valence and conduction band edges at the interface and on the right the energies are shown corresponding to the first ($\ce{DH^-/D^{\cdot-}}$) and second ($\ce{D^{\cdot -}/D}$) oxidation potentials of the two-electron redox species. Reversible electron transfers from $\ce{DH^-}$ to the valence band ($v$) and $\ce{D^{\cdot -}}$ to the conduction band $c$ will contribute to successful bifurcating current, but energy wasting short-circuit reactions are also possible (D) Productive reactions conserve the reducing power of the reactants and contribute to bifurcating current. Short circuit reactions (which can be prevented, \textit{vida infra})  reduce a high-potential acceptor (blue) from a low-potential donor (red) and waste free energy. }
\end{figure}

The electron bifurcating junction described in this paper is illustrated in Figure \ref{figure2}A: a three-way n-p-electrolyte junction. The insulating region prevents unnecessary electron hole recombination deep in the semiconductor. The depletion region extends across the entire semiconductor electrolyte interface due to interface polarization. 

The electrolyte contains a two-electron redox species \ce{DH^-} that is oxidized on a semiconducting surface, injecting charges into a semiconductor bridge separating n- and p-doped regions. The two-electron redox species D  can perform the following one-electron half-reactions:
\begin{equation}
    \ce{D + e^- <->[\Delta G^\circ_{D/D^{\cdot-}}] D^{\cdot-}}
\end{equation}
\begin{equation}
    \ce{D^{\cdot-} + e^- + H^+ <->[\Delta G^\circ_{\ce{D^{\cdot-}}/\ce{DH^-}}] DH^-}.
\end{equation}
The unstable radical intermediate $\ce{D^{\cdot -}}$ must have a much lower oxidation potential than the fully reduced $\ce{RH^-}$. This inverted\cite{evans2008one, yuly2021efficient, arrigoni2020rational} (or ``crossed''\cite{nitschke2012redox, peters2016electron}) ordering of the potentials enables the D species to be a two-electron carrier\cite{yuly2021efficient}, and also gives the radical intermediate enough energy to inject charge into the conduction band of the semiconductor\cite{gerischer1969charge}. In nature, molecules such as flavins, quinones, and NADPH can serve as bifurcating two electron donors\cite{peters2016electron, buckel2018flavin,buckel2018flavin2, sarewicz2021catalytic, yuly2019electron}. Other organic compounds with these properties exist, such as methanol\cite{hykaway1986current}, formate\cite{morrison1967chemical}, and others\cite{kalamaras2015current}. In fact, high energy one-electron intermediates are commonplace in proton-coupled electron transfer reagents\cite{agarwal2021free}.

When $\ce{D^=}$ comes into contact with the semiconductor surface, a sequence of electron transfers  initiates an electron bifurcation event, illustrated in Figure 3B. This sequence is analogous to that shown in Figure \ref{figure1}B for the natural enzymes. First, the fully reduced $\ce{DH^-}$ occasionally encounters a hole \ce{h^+}, which is annihilated as it oxidizes \ce{DH^-} to form the radical $\ce{D^{\cdot-}}$ species. As described above, the $\ce{D^{\cdot-}}$ is highly unstable and thus has sufficient reducing power to inject an electron into the conduction band of the semiconductor, where it preferentially flows to the n-doped region. The second electron does not follow the first electron into the valence band because the hole that accepted the first electron has been filled. The energy levels (one-electron redox potentials) of the \ce{DH^-} species semiconductor bands at the interface are shown in Figure 3B. This sequence of oxidation events at a semiconductor electrode is implicated in the current doubling mechanism in photoelectrochemistry\cite{morrison1967chemical,gerischer1969charge, hykaway1986current,kalamaras2015current}, but here with the addition of the n- and p-doped regions and without any involvement of light.

For electron bifurcation to spontaneously occur, several reactions and recombination processes must be suppressed. Figure \ref{figure2}B shows the energy levels of the two-electron species and valence and conduction bands. The solid red and blue arrows indicate productive electron transfer reactions, and the pink dashed arrows indicate short circuit reactions. Figure \ref{figure2}D lists all relevant reactions, both at the interface (heterogeneous) and in the bulk (homogeneous). All the short-circuit reactions can be adequately suppressed.

\textbf{Suppressing interfacial short circuits}

The radical species $\ce{D^{\cdot -}}$ participates in two interfacial short circuit reactions: $\ce{D^{\cdot-}}$ oxidation by a valence band hole and $\ce{D^{\cdot-}}$ reduction by a conduction band electron (see Figure \ref{figure2}C)
\begin{equation}
\begin{split} \label{Eq:interfaceSC}
    \ce{D^{\cdot-}} + \ce{h}^+ \rightarrow \ce{D},\\
    \ce{D^{\cdot-}} + \ce{e}^- + \ce{H+} \rightarrow \ce{DH^-}.
\end{split}
\end{equation}
To successfully bifurcate current across the junction, these reactions much occur at a much smaller rate than the rates of productive electron transfers
\begin{equation}
\begin{split}
    \ce{D^{\cdot-}} \rightarrow \ce{D} + \ce{e-},\\
    \ce{D^{\cdot-}} + \ce{H+} \rightarrow \ce{DH^-} + \ce{h+}. \\
\end{split}
\end{equation}
Little is required to adequately suppress reactions \ref{Eq:interfaceSC}: the semiconductor must merely be far from degenerate. If this is the case, there are far fewer holes and electrons to catalyze short circuit reactions (Eq \ref{Eq:interfaceSC}) than free states in the valence and conduction bands involved in productive reactions. This constraint is derived quantitatively in the Supporting Information.

\textbf{Suppressing homogeneous short circuits (semiconductor)}

In biological electron bifurcation, successful separation of the charges to $L_1$ and $H_1$ (as shown in Figure \ref{figure1}B), does not itself guarantee successful bifurcation, as such events could be overwhelmed by short-circuiting from the $\ce{A_L}$ pool to the $\ce{A_H}$ pool\cite{yuly2021energy}. Analogously, in a bifurcating semiconductor-electrolyte junction, recombination of electrons and holes in the semiconductor may overwhelm successful charge injection into the valance and conduction bands. Electron hole recombination is defined by the process
\begin{equation} \label{Eq:ehrecomb}
    \ce{e^-} + \ce{h^+} \rightarrow \emptyset.
\end{equation}

Electrons and holes can recombine when they encounter each other (bimolecular recombination \cite{lakhwani2014bimolecular}), or by first being absorbed by intermediate trap states (trap-assisted recombination \cite{kuik2011trap,zeiske2021direct}). To achieve bifurcated current, the current generated by recombination across the junction $I_{\text{Rec}}$ must be smaller than the current generated by productive electron transfer from the electrolyte $I_{\text{inj}}$. If $I_{\text{rec}}<< I_{\text{inj}}$, then bifurcation is guaranteed, as the injected charges have no other option but to bifurcate. Thus, I define the bifurcation efficiency as
\begin{equation}\label{Eq:eff}
    \eta_{\text{eff}} = \left(1 -\frac{I_{\text{rec}}}{I_{\text{inj}}}\right)
\end{equation}
when $\eta_{\text{eff}} = 1$, the injected current is perfectly bifurcated across the junction, half towards the p-side and half-towards the n-side. This definition of efficiency permits negative values, which indicates recombination dominates and no energy transduction occurs at all.

Bimolecular recombination follows a local second-order rate law
\begin{equation}\label{Eq:recrate}
    R_{\text{rec}} = k_{\text{rec}} np,
\end{equation}
where $k_{\text{rec}}$ is the recombination strength. In this paper I assume the worst case: that electrons and holes always recombine when they come into electrostatic range. This is Langevin recombination\cite{burke2015beyond, van2009electron}, with $k_{\text{rec}}  = k_{\text{Langevin}} = |e^-|\mu/\epsilon$. The quantity $\mu = \mu_{h^+} + \mu_{e^-}$ is the sum of electron and hole mobilities, and $\epsilon$ is the permittivity of the semiconductor.
The key insight that to be exploited in bifurcating junctions is that the rate of charge injection into the semiconductor is quasi-first order. When the $c_{\ce{DH-}}/c_{\ce{D}}>>1$ such that the junction is poised in the forward (bifurcating) direction,
\begin{equation}
    R_{\text{inj}} = k_{\ce{DH^-}\rightarrow v} \hspace{2pt} c_{\ce{DH^-}} p
\end{equation}
where the concentration $\ce{DH-}$ is in excess. The rate constant $k_{\ce{DH^-}\rightarrow v}$ is discussed below. Because recombination and heterogeneous electron transfer have different dependencies on the carrier concentrations $n$ and $p$, their relative rates can be set by adjusting the bias voltage $V_{\text{bias}}$. In particular, at low bias, charge injection occurs much faster than recombination.

At high bias, recombination short-circuiting will overwhelm productive charge injection from the electrolyte, and the junction will behave like an ordinary p-n diode. In the ideal diode theory\cite{shockley1949theory}, the current depends on the bias exponentially 
\begin{equation} \label{Eq:Irec}
    I_{\text{rec}} = 2|e^-|\int_V dV \hspace{3pt} k_{\text{rec}}  np \propto \exp \left(\frac{|e^-|V_{\text{bias}}}{\eta k_B T} \right)
\end{equation}
where the integral is over the semiconductor volume $V$, and the exponential dependence occurs when $|e^-|V_{\text{bias}}$ is greater than a few $k_BT$. The ideality factor $\eta = 1$ for an ideal diode where biomolecular recombination dominates. This value of $\eta$ reflects the second-order kinetics of bimolecular recombination (Eq \ref{Eq:recrate}). If trap-assisted (Shockley-Reed-Hall) recombination dominates, the ideality factor is $\eta = 2$ \cite{sah1957carrier}. 

In contrast, the rates of productive reactions at the interface are proportional to $n$ or $p$ separately, so that
\begin{equation} \label{Eq:Iinj}
    I_{\text{inj}} = 2|e^-| \int_S dS \hspace{3 pt}k_{\ce{DH-}\rightarrow v} c_{\ce{DH-}} p \hspace{3 pt}\propto \exp(\frac{|e^-|V_{\text{bias}}}{2 k_B T}).
\end{equation}
The integral is over the surface $S$ of the semiconductor-electrolyte interface. The factor of $2$ in the denominator reflects the pseudo-first order kinetics of charge injection. Thus, if trap-assisted recombination does not completely dominate (i.e. $\eta < 2 $), then as the bias $V_{\text{bias}}$ decreases, the bifurcating efficiency (Equation \ref{Eq:eff}) can be made as high as desired. The exponential dependence of $I_{\text{rec}}$ and $I_{\text{inj}}$ is observed quantitatively in the simulation of the model junction described in the next section.

Besides exploiting the exponential dependence of carrier concentrations on $V_{\text{bias}}$, other strategies are necessary to reduce recombination short-circuiting. In particular, materials with low carrier mobility $\mu$ will have slower recombination (assuming $k_{\text{rec}} = k_{\text{Lengevin}}$), as electron-hole encounters would be less frequent. In simulations of the example bifurcating junction described in below, the mobility was varied over three orders of magnitude in an ultra-low mobility regime ($\mu$ from $ \sim 10^{-3} $ to $\sim 10^1\text{ cm}^2 \text{ V}^{-1} \text{sec}^{-1}$, see Supporting Information).  Disordered organic semiconductors are routinely reported with mobilities in this range\cite{coropceanu2007charge}. Below a value of $\mu \sim 10^{-1} \text{ V}^{-1} \text{sec}^{-1}$, the performance of the junction is roughly constant. Thus, while very low mobility materials are likely desired, fine tuning of mobilities is likely unnecessary below a threshold. Care should be taken to avoid introducing many deep trap states along with the disorder required to achieve low mobility \cite{kuik2011trap, zeiske2021direct}.  

\textbf{Suppressing homogeneous short circuits (electrolyte)}

The final class of short-circuit processes to discuss occurs in the electrolyte, namely any process that oxidizes $\ce{D^{\cdot -}}$ without injecting an electron into the conduction band of the semiconductor. By purifying the electrolyte solution of unwanted oxidants, the lifetime of the $\ce{D^{\cdot -}}$ species may be lengthened, allowing the radicals to inject charge into the conduction band. High-efficiency injection of radical charge into a semiconductor conduction band has been demonstrated many times\cite{morrison1967chemical, gerischer1969charge, hykaway1986current, kalamaras2015current,seshadri1991effect}.

There is one short circuit reaction that seems impossible to prevent by changing the electrolyte composition, namely the disproportionation of the $\ce{D}$ species
\begin{equation}\label{Eq:disproportionation}
    2\ce{D^{\cdot -} + H^+} \rightarrow \hspace{2 pt} \ce{DH^- + D}.
\end{equation}
This reaction is a short circuit process because $\ce{D^\cdot-}$ is both a high-energy electron donor and a low-energy electron acceptor (see the energy levels depicted in Figure \ref{figure2}C). Assuming an excess of protons, the rate of disproportionation is 
\begin{equation}\label{Eq:disprate}
    R_{\text{disp}} = k_{\text{disp}} c_{\ce{D^\cdot-}}^2
\end{equation}
where $k_{\text{disp}}$ is a rate constant and $c_{\ce{D^\cdot-}}$ is the concentration of $\ce{D^\cdot-}$.
This process cannot be completely eliminated because $\ce{D^\cdot-}$ is part of the bifurcation process by injecting its charge into the conduction band (with rate constant $k_{\ce{D^{\cdot-}}\rightarrow c}$). At equilibrium, the concentration of the $\ce{D^{\cdot -}}$ species is small, as the $\ce{D^{\cdot-}}$ radical is high energy. The more energetically unstable the $\ce{D^{\cdot -}}$ radical, the lower its equilibrium concentration. As $\ce{DH-}$ is oxidized by holes at the surface, the concentration of $\ce{D^{\cdot -}}$ radicals will increase, creating a driving force for disproportionation. Fortunately, this process also has a favorable scaling with $V_{\text{bias}}$
\begin{equation}\label{Eq:dispvdepend}
    R_{\text{disp}} \leq K \exp\left(\frac{|e^-|V_{\text{bias}}}{k_BT} \right)  
\end{equation}
where $K$ has no $V_{\text{bias}}$ dependence. Equation \ref{Eq:dispvdepend} is derived in the Supporting Information. Thus, by decreasing $V_{\text{bias}}$, disproportionation short-circuiting will decrease exponentially faster than productive reactions $R_{\text{inj}}$.

\section{Constraints and performance of a model bifurcating junction} \label{section: constraints}

\begin{figure}[H]\centering
\includegraphics[scale=0.50]{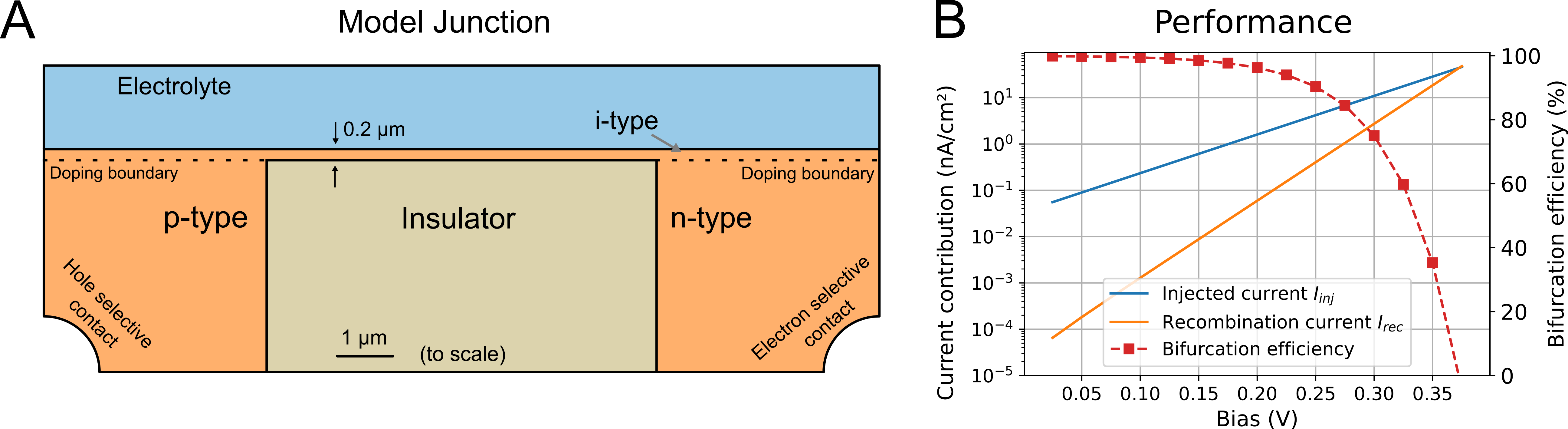}
\caption{\label{figure3} \textbf{Model bifurcating junction and its performance in simulation} (A) Model electron bifurcating junction (to scale). The dimensions of the model were chosen so that the depletion region could be fully accommodated within the device. (B) Simulated performance of the model bifurcating junction. The injected $I_{\text{inj}}/A$ (blue) and recombination $I_{\text{rec}}A$ (orange) currents both increase exponentially with $V_{\text{bias}}$, but with  the predicted different exponential factors (Equations \ref{Eq:Iinj} and \ref{Eq:Irec}). Both currents are divided by the surface area $A$ of the interface for meaningful comparison. Thus, at low currents the bifurcation efficiency $\eta_{\text{eff}}$ (red, in percent) is very high. Short-circuits dramatically reduce efficiency when $V_{\text{bias}} \approx 0.2-0.25$ V or greater, where the bifurcated current is $\sim 1 - 10$ nA/cm$^2$.
}
\end{figure}

To validate the theory of bifurcating junctions as described above, I simulated a model junction with typical material properties and charge-transfer kinetics. The geometry of this junction is illustrated to scale in Figure \ref{figure3}A. The properties of the junction and model have many parameters. As discussed below, I chose values of the parameters that are typical. A table of all parameters used in the simulation is found in the Supporting Information.

Several key parameters describe the kinetics of electron transfer across the semiconductor-electrolyte interface. For optimal performance, the semiconductor surface should not be chemically modified during operation, as this would limit the lifetime and reproducibility of the bifurcating junction. A weak electronic coupling between the semiconductor and the $\ce{D}$ molecule minimizes this risk. This weak coupling pushes the interfacial electron transfers into the electronically non-adiabatic regime\cite{nitzan2024chemical}.

Theories of nonadiabatic heterogeneous electron transfer as discussed by Marcus \cite{marcus1965theory} and others \cite{gerischer1969charge, gerischer1990impact, lewis1998progress, lewis1991analysis} provide a powerful framework for understanding these kinetics. While these theories do not capture every aspect of the dynamics at the interface, the rate constants from these theories are typical and thus useful to establish baseline expectations for the performance of bifurcating junctions. In the framework of these theories, the rate constant for transfer of a semiconducting carrier (electron or hole) to a redox species in an electrolyte phase is\cite{gerischer1969charge, gerischer1990impact, lewis1991analysis, royea1997fermi, lewis1998progress}
\begin{equation} \label{Eq:etrate}
    k_{\text{ET}} = C \exp \left(\frac{(\Delta G^\circ + \lambda)^2}{4\lambda k_B T} \right).
\end{equation}
where $k_B T$ is the thermal energy, $\Delta G^\circ$ is the standard free energy of the electron transfer, and $\lambda$ is the total reorganization energy, and $C$ is a prefactor (see below). There are several different frameworks used to estimate the prefactor $C$  (see references \cite{royea1997fermi} and \cite{gerischer1991electron} for example) but they agree that $C$ cannot reasonably exceed $10^{-17}-10^{-16} \text{cm}^4/\text{sec}$ for nonadiabatic electron transfer between a semiconductor and a freely diffusing redox species. The rate constant of the reverse process (charge injection into the semiconductor that creates an electron or hole species), can be derived from Equation \ref{Eq:etrate} using detailed balance\cite{shreve1995analytical}
\begin{equation}\label{Eq:detailedbalance}
    k^r_{\text{ET}} = k_{\text{ET}} \hspace{2 pt} e^{-\Delta G^\circ/k_BT} = k_{\text{ET}} N_{c,v} \exp\left(\frac{E^\circ(A/A^{\pm}) - E_{v,c}}{k_BT} \right),
\end{equation}
where $N_{c,v}$ is the effective density of states in the conduction or valence band (whichever is losing a charge) and $E^\circ(A/A^-)$ is the standard free energy for redox species A to gain an electron, that can be calculated from a standard reduction potential of A\cite{reiss1985absolute}. Thus, the electron transfers between $\ce{D}$ and the semiconductor will be rate limiting for the bifurcating current in a bifurcating junction. If the junction is designed to operate equally well in the forward (bifurcating) and reverse (confurcating) directions, then $E^\circ(\ce{D}/\ce{D^\cdot-}) = E_c $ and $E^\circ(\ce{D^{\cdot-/}}\ce{DH^-}) = E_v$ so that the energy levels of the redox species are lined up with the valance and conduction band edges. A reorganization energy $\lambda \approx 0.5 $ eV is possible \cite{lewis1991analysis}, but significantly smaller reorganization energies might be difficult. Assuming $\lambda \approx 0.5 $ eV, the fastest rate constant for the $\ce{RH^-} \rightarrow v$ and $c \rightarrow \ce{R^{\cdot -}}$ that can be reasonably expected (without surface adsorption) is $\approx 10^{-19}-10^{-18} \text{ cm}^4/\text{sec}$ (Eq \ref{Eq:etrate} and Ref \cite{lewis1998progress}).

Material properties for simulating the model junction were chosen to reflect typical values for low-mobility semiconductor electrodes. The intrinsic carrier density in the semiconductor was assumed to be $n_i = N_{v,c} e^{-E_b/2k_bT}$, where $N_{v,c} = 10^{20}/\text{cm}^3$ is the effective density of states (assumed to be the same in the valence and conduction bands) and and $E_b = 1.6 $ eV is the band gap (chosen to be achievable using organic low-mobility materials but not too wide as to make doping difficult \cite{ganose2022defect}). This sets a length scale for the depletion region, which was sets the required scale for the volume of the semiconductor (i.e. the depth of the semiconductor must be $\sim 5$ $\mu$m). Critically, this sets the volume $V$ of the semiconductor region, which should be as small as possible to reduce recombination (Equation \ref{Eq:Irec}, this is discussed further in the next section). The dopant concentrations were set to establish a built-in potential $V_{\text{bi}} = 0.9 $ V. The bias $V_{\text{app}}$ is applied using a hole-selective contact on the p-doped side and an electron-selective contact on the n-doped side. The carrier mobility for both carriers was set to $\sim 10^{-1} \hspace{3pt }\text{cm}^2 \hspace{2 pt} \text{V}^{-1} \text{sec}^{-1}$ for the simulation in Figure 3B. The simulation was performed over a range of mobilities $\mu$, with little effect on the performance provided $\mu < 10^{-1} \text{V}^{-1} \text{sec}^{-1}$ (see Supporting Information).

The electrolyte solution was chosen to have a relative permittivity of $\epsilon^{e}_r = 20$, typical of polar solvents, and with redox-inactive positively and negatively charged ions $c_{\text{ion}} \approx 5.7 \times 10^{15}\text{ cm}^{-3}$ to achieve a screening Debye layer length of $L_D = 50$ nm. A large concentration of the redox molecule $\ce{D}$ is desirable to maximize the opportunities to inject charges into the semiconductor to bifurcate.  I assume a concentration of $c_{\ce{DH^-}}, c_{\ce{D}} \approx 1$ M, although somewhat higher concentrations could be achieved if the solvent was pure $\ce{DH-}$.

Using the COMSOL Multiphysics\textsuperscript{®}software\cite{comsol}, the semiconductor drift-diffusion equations were coupled with the Poisson-Boltzmann equations (for the electrolyte domain) and solved at steady-state under an applied bias $V_{\text{bias}}$. I assumed that the dominant short-circuiting mechanism is Lengevin bimolecular recombination in the semiconductor. The interfacial short-circuits and disproportionation short-circuits are much more easily prevented (see above). The simulation is described in more detail in the Supporting Information.

The results of the simulation are shown in Figure 3B. Exponential dependence of the injected current $I_{\text{inj}}$ and recombination current $I_{\text{rec}}$ are as predicted by the theory described in the previous section. Thus, very high bifurcation efficiencies can be achieved at low bias $V_{\text{bias}}$, and efficiency dramatically drops when the bias is greater than $\sim 0.20- 0.25 $ V, where the bifurcated current reaches a maximum value of $\sim 1- 10 $ nA/cm$^2$. The next section discusses strategies for more advanced junctions with higher performance.

\section{Looking forward: applications and enhancing performance}

The junction design in Figure 3A was designed to be illustrative, not optimal. Future designs would likely be greatly improved by a single factor: using layered semiconductor heterojunctions instead of a homogeneous np junction. When the semiconductor domain semiconductor is made of one (inhomogeously doped) material, then the bandgap, built-in potential, and widths of the depletion region at the interface cannot all be chosen independently\cite{sze2021physics}. This constrains the volume of semiconductor.
To minimize short-circuiting, the volume of the semiconducting region where recombination occurs should be minimized, as productive reactions only occur at the electrolyte interface. By layering semiconductors with different bandgaps and electron affinities, the energy landscape for electrons and holes within the junction could be sculpted without relying on dopants. This would allow for much narrower semiconductor electrodes to be used. An analogous development took place in semiconductor light harvesting. A homogeneous np junction is capable of harvesting light\cite{sze2021physics}, but modern photovoltaic cells use layered heterojunctions to achieve depletion regions with smaller sizes and other favorable properties.

It is also possible that bulk heterojunctions might be used at the electrolyte surface, the electron donor material serving as a source of holes and the electron accepting material serving as the source of electrons. Such a design is intriguing because many bulk-heterojunction organic solar cells have been observed to have $k_{\text{rec}}$ much smaller than $k_{\text{Langevin}}$ \cite{lakhwani2014bimolecular} (by orders-of-magnitude). One explanation for such small recombination rate constants is that encounters between electrons and holes at heterojunctions do not always result in recombination, instead forming quasi-stable charge transfer states\cite{burke2015beyond}. If bifurcating junctions were built from these materials, higher carrier densities (and therefore larger bifurcated currents) might be possible without increasing recombination short-circuiting.   

Other paths to increased performance are possible. For bifurcating semiconductor electrodes, the interfacial electron transfers are rate limiting (unlike metal electrodes that are diffusion limited \cite{menshykau2008influence}), so increasing effective surface area (by increasing roughness) might boost the bifurcated current. Lastly, careful design of the proton-coupled electron transfer kinetics at the semiconductor surface\cite{warburton2022theoretical} might further increase bifurcating current densities.

The bioinspired bifurcating junctions described here may inspire several possible applications. For example, electron bifurcating junctions might serve as efficient direct current voltage converters. Placing bifurcating junctions at the terminus of a battery cell would boost the open-circuit voltage of the cell. Similarly, the open-circuit voltage of solar cells could be increased by placing bifurcating junctions downstream of the light harvesting process. Ultimately, bifurcating junctions (perhaps of smaller size than the junction in Figure 3A) could be used to drive catalysis by channeling the high energy electrons from the bifurcation process to desired catalytic sites, as occurs in natural electron bifurcation. The semiconducting material used to bifurcate the electrons might even be used to separate different electrolyte phases, allowing bifurcated electrons from one electrolyte to drive catalysis in another. This separation could enable electrochemistry that is difficult in a single solvent (such as driving reduction outside the redox w redox window). Finally, realizing bioinspired bifurcating systems might allow more meaningful reflection on biological electron bifurcation, raising new questions about the function of nature's elegant molecular machines.

\section{Acknowledgments}
The author thanks Joshua Atkinson, Trevor GrandPre, Rich Pang, and Peng Zhang but especially David Beratan, Andrew Bocarsly, Alejandro Martinez-Calvo, Colin Scheibner, Ned Wingreen, and Qiwei Yu for helpful discussions. This work was funded by the Peter B. Lewis Lewis-Sigler Institute/Genomics Fund through the Lewis-Sigler Institute for Integrative Genomics at Princeton University.

\bibliographystyle{pnas.bst}
\bibliography{bibliography}

%\end{multicols}
\end{document}

\title{Supporting Information for:\\
Designing semiconductor-electrochemical junctions for bioinspired energy transduction}

\author[1]{Jonathon L. Yuly\footnote{email: \href{mailto:name@name.com}{jonathon.yuly@princeton.edu}}}
\affil[1]{Lewis-Sigler Institute for Integrative Genomics, Princeton University, Princeton, NJ, 08540}

\maketitle

\section{Suppressing interfacial short-circuits}\label{Sec:interfacesc}

Because both of the interfacial short circuits require encounters between $\ce{D^{\cdot -}}$ and either an electron or hole, the rates of both of these processes are governed by second-order rate laws\cite{gerischer1969charge, lewis1991analysis, gerischer1990impact, lewis1998progress}
\begin{equation}\label{Eq:SC}
\begin{split}
    R_1 &= k_{\ce{D^{\cdot -}}\rightarrow v} \hspace{2 pt} p \hspace{2 pt} c_{\ce{D^{\cdot -}}}\\
    R_2 & = k_{c \rightarrow \ce{D^{\cdot -}}} \hspace{2 pt} n \hspace{2 pt} c_{\ce{D^{\cdot -}}}
\end{split}
\end{equation}
where $k_{\ce{D^{\cdot -}\rightarrow v}}$ is the rate constant for electron transfer from $\ce{D^{\cdot -}}$ to the valence band and $k_{c \rightarrow \ce{D^{\cdot -}}}$ is for the electron transfer from the conduction band to $\ce{D^{\cdot -}}$. The quantities $n$, $p$, $ c_{\ce{D^{\cdot -}}}$ are the electron, hole, and $\ce{D^{\cdot -}}$ concentrations at the interface. The dependence of $k_{c \rightarrow \ce{D^{\cdot -}}}$ on the local concentration of protons can be neglected by assuming that protons are in excess.

The radical species $\ce{D^{\cdot -}}$ also participates in productive electron transfer, injecting a charge into the conduction band. The rate law for this process is first-order because plenty of states are available in the conduction band to accept an electron, so no encounter between $\ce{D^{\cdot -}}$ and a semiconductor carrier is necessary. Thus, the turnover for productive charge injection into the conduction band is
\begin{equation}\label{Eq:prod}
    \begin{split}
        R^c_{\text{prod}} = k_{\ce{D^{\cdot -}}\rightarrow c} \hspace{2 pt} c_{\ce{D^{\cdot -}}}\\
        R^v_{\text{prod}} = k_{v \rightarrow \ce{D^{\cdot -}}}\hspace{2 pt} c_{\ce{D^{\cdot -}}},
    \end{split}
\end{equation}
where $k_{\ce{D^{\cdot -}}\rightarrow c}$ and $k_{v \rightarrow \ce{D^{\cdot -}}}$ are first-order rate constants. In order for the productive turnover to dominate over short circuits, $R_1, R_2 << R^c_{\text{prod}}, R^v_{\text{prod}}$. Thus, to sufficiently suppress heterogeneous short-circuit reactions, bifurcating junctions as described above must satisfy
\begin{equation} \label{Eq:bif_cond}
\begin{split}
    k_{\ce{
    D^{\cdot -}}\rightarrow v} \hspace{2 pt} p &<< k_{\ce{D^{\cdot -}} \rightarrow c},\\
    k_{c \rightarrow \ce{D^{\cdot -}}} \hspace{2 pt} n &<< k_{\ce{D^{\cdot -}}\rightarrow c}.
\end{split} 
\end{equation}
As described in the main text, $k_{c \rightarrow \ce{D}} = k_{\ce{DH^-} \rightarrow v} \sim 10^{-19} - 10^{-18} \text{ cm}^3/\text{sec}$, and also I assume that the redox energy levels of $\ce{D}$ are lined up with the valence and conduction band edges (that is, $E^\circ(\ce{D}/\ce{D^\cdot-}) = E_c $ and $E^\circ(\ce{D^{\cdot-}}/\ce{DH^-}) = E_v$)  Thus, we use the detailed balance expression \cite{shreve1995analytical}(Eq 14 in the main text) to estimate the rate constants for the productive reactions involving the  $\ce{D^{\cdot -}}$ species, i.e.,
$k_{\ce{D^{\cdot -}} \rightarrow c} = N_c \hspace{2 pt} k_{c \rightarrow \ce{D}} $ and $ k_{v \rightarrow \ce{D^{\cdot -}}} = N_v \hspace{2 pt} k_{\ce{DH^-} \rightarrow v}$. For the interfacial short-circuits, we assume the worst case, that is, activationless electron transfer $\Delta G \approx -\lambda$, with the maximal value of $k_{\ce{D^{\cdot -}}\rightarrow v} = k_{c \rightarrow \ce{D^{\cdot -}}} \approx 10^{-17}-10^{-16} \text{ cm}^3/\text{sec}$. Thus, to satisfy Equations \ref{Eq:bif_cond}, $n$ and $p$ at the semiconductor surface must satisfy
\begin{equation} \label{Eq:bif_cond2}
    n,p << 10^{-2} N_{c,v}.
\end{equation}
This is barely constraining and only requires that the semiconductor be far from degenerate.

\section{Suppressing disproportionation}\label{Sec:disprop}
Disproportionation of $\ce{D}$ is defined by the process
\begin{equation}\label{Eq:disproportionation}
    2\ce{D^{\cdot -} + H^+} \rightarrow \hspace{2 pt} \ce{DH^- + D}.
\end{equation}
At equilibrium, the concentration of $\ce{D^{\cdot-}}$ is very small, given by\cite{michaelis1938theory}
\begin{equation}
    c_{\ce{D^{\cdot-}}}^{2} = c_{\ce{DH^-}} c_{\ce{D}} \exp\left(-\frac{E^\circ(\ce{DH^-}/\ce{D^{\cdot-}}) - E^\circ(\ce{D^{\cdot-}}/\ce{D})}{k_BT} \right),
\end{equation}
where $E^\circ(\ce{DH^-}/\ce{D^{\cdot-}})$ is the energy for the half reaction of the first oxidation $\ce{DH^-}\rightarrow \ce{D^{\cdot -}}$ and $E^\circ(\ce{D^{\cdot-}}/\ce{D})$ is the energy of the half reaction for the second oxidation $\ce{D^{\cdot -}}\rightarrow \ce{D}$. When the $\ce{D^\cdot-}$ radical is energetically unstable, the reduction potentials corresponding to these energies are said to be inverted \cite{evans2008one} (or ``crossed'' \cite{nitschke2012redox}). The more inverted the potentials, the more unstable the intermediate.
When the junction is driven to bifurcate (confurcate), these redox couples will be driven out of equilibrium. The concentration of $D^{\cdot -}$ will increase at the interface as either $\ce{DH^-}$ is oxidized ($\ce{D}$ is reduced) by the valence (conduction) band. The newly formed $D^{\cdot -}$ will either (1) short-circuit via the interfacial circuits described in the main text, (2) diffuse away from the surface, or (3) participate in productive electron transfer at the interface. Section \ref{Sec:interfacesc} above demonstrated how (1) can be suppressed, but the freely diffusing $D^{\cdot -}$ radicals will occasionally encounter each other, creating the opportunity to short-circuit by Equation \ref{Eq:disproportionation}. We assume that the rate of disproportion is given by
\begin{equation}\label{Eq:disprate}
    R_{\text{disp}} = k_{\text{disp}} c_{\ce{D^{\cdot -}}}^2
\end{equation}
where $k_{\text{disp}}$ is the disproportion rate constant, that is often very fast, even diffusion limited\cite{bielski1981mechanism}.

When the bifurcating junction is operating in the forward (bifurcating) regime, the rate of oxidation of $\ce{DH^{\cdot -}}$ is approximately $R_{\text{init}} \approx k_{\ce{DH^-} \rightarrow v} \hspace{2 pt} c_{DH^-} p$. This is the rate of oxidation per surface area of the semiconducting surface to the solution. Assuming the electrolyte phase is well-mixed (little volume seems needed), the total rate of $\ce{DH^-}$ oxidation per volume of electrolyte is $R_{\text{init}}(A/V)$, where $A$ is the surface area of exposed semiconductor and $V_e$ is the total volume of electrolyte. The corresponding rate of (productive) $\ce{D^{\cdot -}}$ oxidation is $\approx k_{\ce{D^{\cdot -}}\rightarrow c}\hspace{2 pt} c_{\ce{D^{\cdot -}}} (A/V_e)$, and the disproportionation rate is $k_{\text{disp}} c_{\ce{D^{\cdot -}}}^2$. Thus, the concentration of the $\ce{D^{\cdot -}}$ radical will change as
\begin{equation}\label{Eq:radrate}
    \frac{dc_{\ce{D^{\cdot -}}}}{dt} \approx  \left( k_{\ce{DH^-}\rightarrow v} \hspace{2 pt} c_{\ce{DH^-}} \hspace{2 pt} p   \hspace{3 pt}- \hspace{3 pt} k_{\ce{D^{\cdot -}}\rightarrow c}\hspace{2 pt} c_{\ce{D^{\cdot -}}}\right)  \frac{A}{V_e} - k_{\text{disp}} c_{\ce{D^{\cdot -}}}^2.
\end{equation}
When the above equation reaches steady state, the concentration $c_{\ce{D^{\cdot -}}}$ is never greater than the upper bound
\begin{equation}\label{Eq:radsteadystate}
    c_{\ce{D^{\cdot -}}} \leq \frac{k_{\ce{DH^-}\rightarrow v} }{k_{\ce{D^{\cdot -}}\rightarrow c}} \hspace{2 pt} c_{\ce{DH^-}} \hspace{2 pt} p.
\end{equation}
Assuming that the valence band is well described with a single quasi-Fermi level $E^L_{f}$ in equilibrium with the left metal contact 
\begin{equation}\label{Eq:boltzhole}
    p \approx n_i \exp\left(\frac{E_v - E^L_f}{k_BT} \right)
\end{equation}
where $E_v$ is the valence band edge. Assuming that the bias $V_{\text{bias}}$ and doping is applied equally on both sides (setting the zero of energy at the middle of the bandgap), then the hole quasi-Fermi level is
\begin{equation}\label{Eq:holequasif}
    E^L_f = -|e^-|\frac{\phi_{\text{bi}} + V_{\text{bias}}}{2} 
\end{equation}
Substituting Equations \ref{Eq:holequasif}, 
\ref{Eq:boltzhole}, and \ref{Eq:radsteadystate} into Equation \ref{Eq:disprate} and simplifying yields
\begin{equation}
    R_{\text{disp}} \leq K \exp\left(\frac{|e^-|V_{\text{bias}}}{k_BT} \right)  
\end{equation}
where $K$ has no $V_{\text{bias}}$ dependence:
\begin{equation}
    K = \left( \frac{k_{\ce{DH^-}\rightarrow v} }{k_{\ce{D^{\cdot -}}\rightarrow c}} \right)^2 k_{\text{disp}}^2 \hspace{2 pt} c_{\ce{DH^-}}^2 \hspace{2 pt} n_i^2 \exp\left(\frac{2E_v + |e^-|\phi_{\text{bi}}}{k_BT}\right)
\end{equation}

Thus, just as the recombination short-circuits at low-enough $V_{\text{bias}}$, the disproportionation short-circuit can always be made smaller than $R_{\text{inj}}$. Furthermore, the disproportionation short circuit can also be sufficiently suppressed relative to productive turnover lowering the concentration of the redox active molecule $\ce{DH^-}$, because $R_{\text{disp}}\propto c_{\ce{D^{\cdot-}}}^2$ but $R_{\text{inj}} \propto c_{\ce{D^{\cdot -}}}$, although this will also slow productive turnover. Furthermore, the geometric factor $A/V_e$ also implies that a large area-to-volume ratio will allow a larger concentration of $\ce{DH^-}$ without significant disproportionation short-circuiting. Thus, in the simulations on the bifurcating junction of Figure 3A, I assume that the disproportionation short-circuiting is negligible. This is because because only a little electrolyte seems needed and disproportionation only becomes relevant when $V_e/A$ becomes large (Equation \ref{Eq:radrate}).

\section{Model electron-bifurcating junction}

\subsection{Parameters}
Table \ref{tab:params} below lists the values of all parameters used to estimate the current through the example bifurcating junction shown in Figure 3 of the main text. For ease of comparison with the equations in the next section, all parameters are listed including those already listed in Table 1 of the main text.

\begin{table}[h]
    \centering
    \begin{tabular}{|c|c|c|c|}
        \hline
         Symbol & Quantity & Value & Reference \\ \hline
         C &  Electron transfer prefactor  & $\sim 10^{-17}  \text{ cm}^4/\text{sec}$  & \cite{lewis1998progress} \\ \hline
         $\lambda$ & Reorganization energy  & $0.5$ eV& \\ \hline
          $k_{\ce{DH^-} \rightarrow v} $ & Productive rate constant    &  $\sim 10^{-19} \text{cm}^4/\text{sec}$   & Main text \\ 
        $k_{c\rightarrow \ce{D}}$& (electron or hole as reactant) & & \cite{lewis1998progress} \\ \hline 
          $k_{v \rightarrow \ce{D^{\cdot-}}}$ & Productive rate constants   & $\sim 10^{-3} \text{ cm}/\text{sec}$   & Main text \\ 
         $k_{\ce{D^{\cdot -}}\rightarrow c}$ & (electron or hole as product) &  & \cite{shreve1995analytical} \\ \hline
        $\mu_s$ &  Carrier mobility & $\sim 10^{-3}-10^0 \text{ cm}^2\text{ cm}^2\text{V}^{-1}\text{sec}^{-1} $ & \cite{kohler2015electronic} \cite{coropceanu2007charge}\\ 
        & & (low mobility materials) &
        \\ \hline
         $E_b$ & Bandgap & 1.6 eV &
        \\ \hline
        $N_{c,v}$ & Effective density of states & $\sim 10^{20}/\text{cm}^3$  & \cite{kohler2015electronic}\\ \hline
         $n_i$& Intrinsic concentration & $ N_{c,v}e^{-E_b/2k_BT} \sim 10^5/\text{cm}^3$$ $ & \cite{kohler2015electronic}
        \\ \hline 
        $\phi_{\text{bi}}$ & Built-in potential of np junction & 0.9 eV &
        \\ \hline
        $N_D$ & Dopant concentration & $\sim 10^{12} \text{ cm}^{-3}$ & Equation \ref{Eq:dopconcent} \cite{selberherr2012analysis}
        \\ \hline
        $\varepsilon^s_r$ & Relative permittivity (semiconductor) & $\approx 5$ &  \\ \hline
        $k_{\text{rec}}$  & Bimolecular recombination   &  $\sim 10^{-9} \text{ cm}^3/\text{sec}$  & $k_{\text{Langevin}}$ \cite{van2009electron}  \\ \hline
        $\epsilon^e_r$ & Relative permitivity (electrolyte) & $\approx 20$ & \\
        \hline
        $L_D$ & Debye length (electrolyte) & $50$ nm & Section  \ref{Sec:electrolyte} \\
        \hline
        $c^0_{+},c^0_-$  & Redox inactive ion bulk concentration & $\sim 10$ $\mu$M & Equation \ref{Eq:Debye}\\
        \hline
        $d$  & Total semiconductor depth & $4$ $\mu$m &   \\ \hline
        $c_{\ce{DH^-}}$  & $\ce{DH^-}$ concentration  & $\sim 1$ M &  \\ \hline
        $d_{\text{bridge}}$ & Bridge thickness & $\sim 0.1 \hspace{2 pt}\mu$m (insensitive) & Simulation \\ \hline
    
    \end{tabular}
    \caption{List of parameters used in simulations of the model bifurcating junction. Rate constants were calculated using the lower range of $C \sim 10^{-17} \text{ cm}^4/\text{sec}$\cite{lewis1998progress}.}
    \label{tab:params}
\end{table}

\subsection{Equations and boundary conditions}

\subsubsection{Semiconductor}
The dimensions of the model junction are shown in Figure \ref{figureS1}. To approximate the carrier distributions $n$ (electrons) and $p$ (holes) inside the junction, the semiconductor equations were used to model the charge carriers in the semiconductor. The electron and hole currents are\cite{Ashcroft}
\begin{equation}
    \mathbf{J}_n = - \mu n \mathbf{E} - D \nabla n  
\end{equation}
\begin{equation}
    \mathbf{J}_p = \mu p\mathbf{E} - D\nabla p
\end{equation}
where $\mathbf{E} = - \nabla \phi$ is the electric field and the electron and hole mobilities are assumed to be the same ($\mu_s \sim 10^{-2} \text{cm}^2/\text{V}\cdot \text{sec} $), and the electron and hole diffusion constants satisfy the Einstein relation
\begin{equation}
   D_s = \mu_s \frac{k_B T}{|e^-|}.
\end{equation}
The potential $\phi$ satisfies the Poisson equation
\begin{equation} \label{Eq:Poisson}
    \nabla \cdot \left(\nabla \epsilon \phi \right) = \rho = |e^-|(p -n + N_D),
\end{equation}
where $N_D$ is the dopant concentration,  distributed uniformly $N_D \approx 10^{14}/\text{cm}^3$ in the n- and p-doped regions, with smooth transition region of width $0.1$ $\mu$m at the doping boundaries. The dopant concentration determines the built-in potential $\phi_{\text{bi}}$ through the equation\cite{selberherr2012analysis}
\begin{equation}\label{Eq:dopconcent}
    N_D = -2 n_i \sinh \left(\frac{|e^-|\phi_{\text{bi}}}{2k_BT} \right)
\end{equation}

Inside the semiconductor domains, the carriers are locally conserved except for recombination and emission processes
\begin{equation}\label{Eq:electroncons}
    \frac{\partial n}{\partial t} = \nabla \cdot \mathbf{J}_n - R_{\text{rec}} + R_{\text{em}}
\end{equation}
\begin{equation}\label{Eq:holecons}
    \frac{\partial p}{\partial t} = \nabla \cdot\mathbf{J}_p - R_{\text{rec}} + R_{\text{em}}
\end{equation}
where $R_{\text{rec}}$ is the recombination rate and $R_{\text{em}}$ is the emission rate. In this simulation, all recombination was assumed bimolecular such that $R_{\text{rec}} = k_{\text{rec}} np$ and the emission rate was set to satisfy detailed balance \cite{shreve_analytical_1995}
\begin{equation}
    R_{\text{em}} = k_{\text{em}} = n_i^2 k_{\text{rec}} 
\end{equation}
where $n_i = N_{c,v}\exp\left(-E_b/2k_bT \right)$ is the intrinsic carrier density.
In the ideal case, the electron- and hole- selective metal contacts are perfectly selective. At these contacts, the majority carrier is assumed to achieve approximate neutrality 
\begin{equation}
    \max \{n, p\} \approx |N_D|/|e^-| 
\end{equation}
where the absolute value $|N_D|$ is taken because $N_D$ has different signs (charge) on the n- and p-doped sides. 
The built-in potential was set to $\phi_{\text{bi}} = 0.9$ eV. At the selective contacts, the minority carriers are blocked from flowing through the contact
\begin{equation}\label{Eq:selective}
    \mathbf{J}_{\min \{ n, p\}} \cdot \hat{n} = 0.
\end{equation}
No contact is perfectly selective, so equation \ref{Eq:selective} is an idealization. Lastly, the potential $\phi$ at the left (L) and right (R) contacts the potential satisfies\cite{selberherr2012analysis}
\begin{equation}
    \phi_{R,L} = \pm \frac{ V_{\text{bias}} +\phi_{\text{bi}} }{2} 
\end{equation}
where $+$ is for R, $-$ is for L, and $V_{\text{bias}}$ is the applied bias.

At all other boundaries, the electron and hole carriers $n$ and $p$ were set to be reflective
\begin{equation}
    \mathbf{J}_n \cdot \hat{n} = \mathbf{J}_p \cdot \hat{n} = 0,
\end{equation}

The relative permittivity of the semiconducting regions and the insulating regions was assumed to be $\epsilon^s_r \approx 5$, and was assumed to contain no free charges. 

For all remaining boundaries ``artificial'' \cite{selberherr2012analysis} boundary conditions were assumed.
\begin{equation} \label{Eq:artificialbc1}
    \nabla \phi \cdot \hat{n}  = 0,
\end{equation}
\begin{equation}\label{Eq:artificialbc2}
    \mathbf{J}_{n} \cdot \hat{n} = 0,
\end{equation}
\begin{equation} \label{Eq:artificialbc3}
    \mathbf{J}_{p} \cdot \hat{n} = 0.
\end{equation}
For the ``artificial'' boundaries on the right and left this boundary condition is justified by symmetry when the system is tiled (other bifurcating junctions with the same dimensions are adjacent). Fort the other boundaries, these conditions can be justified when large insulating regions exist on the other side\cite{selberherr2012analysis}.

\subsubsection{Electrolyte} \label{Sec:electrolyte}

The electrolyte solution is assumed to contain positively ($+|e^-|)$ and negatively ($-|e^-|$) charged species (with bulk concentrations $c^0_+$, and $c^0_-$, respectively) that do not interact chemically with the semiconductor and set the ionic strength of the solution to a value such that the Debye length of the solution is
\begin{equation}\label{Eq:Debye}
    L_D = \sqrt{\frac{\varepsilon^e_r \varepsilon_0 k_B T}{|e^-|^2\left(c^0_- + c^0_+\right)}} \approx 50 \text{ nm}.
\end{equation}
This length was chosen so that the ionic layers would be clearly visible in Figure \ref{figureS2}. Choosing a smaller Debye length would not significantly change the results, because the potential at the interface is already not significantly different from $0$ V (see Figure \ref{figureS2}), and is uniform across the interface. The effects of any Stern layer are neglected, but would make the potential even more uniform across solvent volume that is accessible by the $D$ molecule. Thus, concentrations of $c_{\ce{DH^-}}$,$c_{\ce{D^\cdot -}}$ , and $c_{\ce{D}}$ would be roughly equivalent across the interface assuming sufficient mixing. The concentrations of $H^+$ and charged forms of the $D$ molecule are assumed to be small compared with the ionic strength generated by the redox inactive ions. If the solvent is itself $\ce{DH-}$, recall that $\ce{DH-}$ only labels a redox state of $\ce{D}$, and does not define the net charge (the oxidized $D$ may be positively charged). 

Using the COMSOL® transport of charged species interface\cite{comsol}, the redox-inactive positive and negatively charged ions were defined to satisfy the drift-diffusion equations
\begin{equation}
    \frac{\partial c_+}{\partial t} = - \nabla \cdot \mathbf{J}_{c_+} = - \nabla \cdot \left(\mu_e c_+ \mathbf{E} - D_e\nabla c_+ \right)
\end{equation}
\begin{equation}
    \frac{\partial c_-}{\partial t} = - \nabla \cdot \mathbf{J}_{c_{-}} = - \nabla \cdot\left(- \mu_e c_- \mathbf{E} - D_e\nabla c_+ \right).
\end{equation}
along with the Poisson equation (Equation) which is global over all domains. The diffusion constant of the electrolyte was set to $D_e = 1.5 \times 10^{-5}\text{cm}^2/\text{sec}$ and $\mu_e = D_e |e^-|/k_B T$ but this quantity does not influence the  steady-state solution, which is equivalent to solving the Poisson-Boltzmann equation
\begin{equation}
   \nabla \cdot \left(\nabla \epsilon \phi \right) = \sum_{i = \pm} c^0_i q_i \exp\left(-\frac{q_i \phi}{k_B T}\right)
\end{equation}
with $q_\pm = \pm |e^-|$ and the bulk concentrations $c^0_+$ and $c^0_-$ satisfy Equation \ref{Eq:Debye}.
For the top, left, and right electrolyte boundaries, ``artificial" boundary conditions similar to Equations \ref{Eq:artificialbc1}, \ref{Eq:artificialbc2}, and \ref{Eq:artificialbc3} were used:
\begin{equation}
    \nabla \phi \cdot \hat{n}  = 0,
\end{equation}
\begin{equation}\label{Eq:ionbound1}
    \mathbf{J}_{c_+} \cdot \hat{n} = 0,
\end{equation}
\begin{equation} \label{Eq:ionbound2}
    \mathbf{J}_{c_-} \cdot \hat{n} = 0.
\end{equation}

\subsubsection{Boundary conditions at the semiconductor-electrolyte interface}

At the electrolyte-semiconductor interface, currents are generated by inhomogeneous charge transfer initiated by oxidation of the $\ce{DH-}$ molecule. This molecule is assumed greatly in excess and also that $c_{\ce{DH-}}>>c_{\ce{D}}$ (the bifurcation process is strongly driven in the forward direction). As discussed in the previous section, I assumed (based on the calculated potential profiles and a well-mixed assumption) that that the concentrations of $c_{\ce{DH^-}}$,$c_{\ce{D^\cdot -}}$ , and $c_{\ce{D}}$ would be roughly equivalent across the interface. These assumptions seem adequate estimate target device performance.

The charge injection from the electrolyte was calculated using Equation 13 from the main text, such that
\begin{equation} \label{Eq:pelectrolyteboundary}
    \mathbf{J}_p \cdot \hat{n} = R_{\text{init}} = k_{\ce{DH^-} \rightarrow v} \hspace{3 pt} p \hspace{3 pt} c_{\ce{DH^-}}.
\end{equation}
Assuming short circuiting in the electrolyte is slow (see section \ref{Sec:disprop}), and there are no other surfaces or oxidants absorbing electrons from $\ce{D^{\cdot -}}$, then almost all the radicals will eventually deposit their electrons into the conduction band. Thus, at steady state
\begin{equation}\label{Eq:nelectrolyteboundary}
    \mathbf{J}_n \cdot \hat{n} \approx -R_{\text{init}}.
\end{equation}
Equations \ref{Eq:pelectrolyteboundary} and \ref{Eq:nelectrolyteboundary} were used in the simulation of the model bifurating junction.
%In the simulation, the concentration of $c_{\ce{DH^-}}$ is taken to be large enough such that the semiconductor region can reach steady state conditions without significantly changing $c_{\ce{DH^-}}$. Using the values of the parameters given in Table \ref{tab:params} and the scale of the model junction
%\begin{equation}
%    R_{\text{crit}}=\frac{k^2_{\ce{D^{\cdot -}}\rightarrow c}}{ k_{\text{disp}}}\hspace{2 pt} \frac{A}{V_e} \approx 10^{14} \text{ sec}^{-1}
%\end{equation}

\subsubsection{Mesh, numerical methods, and resolved boundary layers}

\begin{figure}[H]\centering
\includegraphics[scale=0.75]{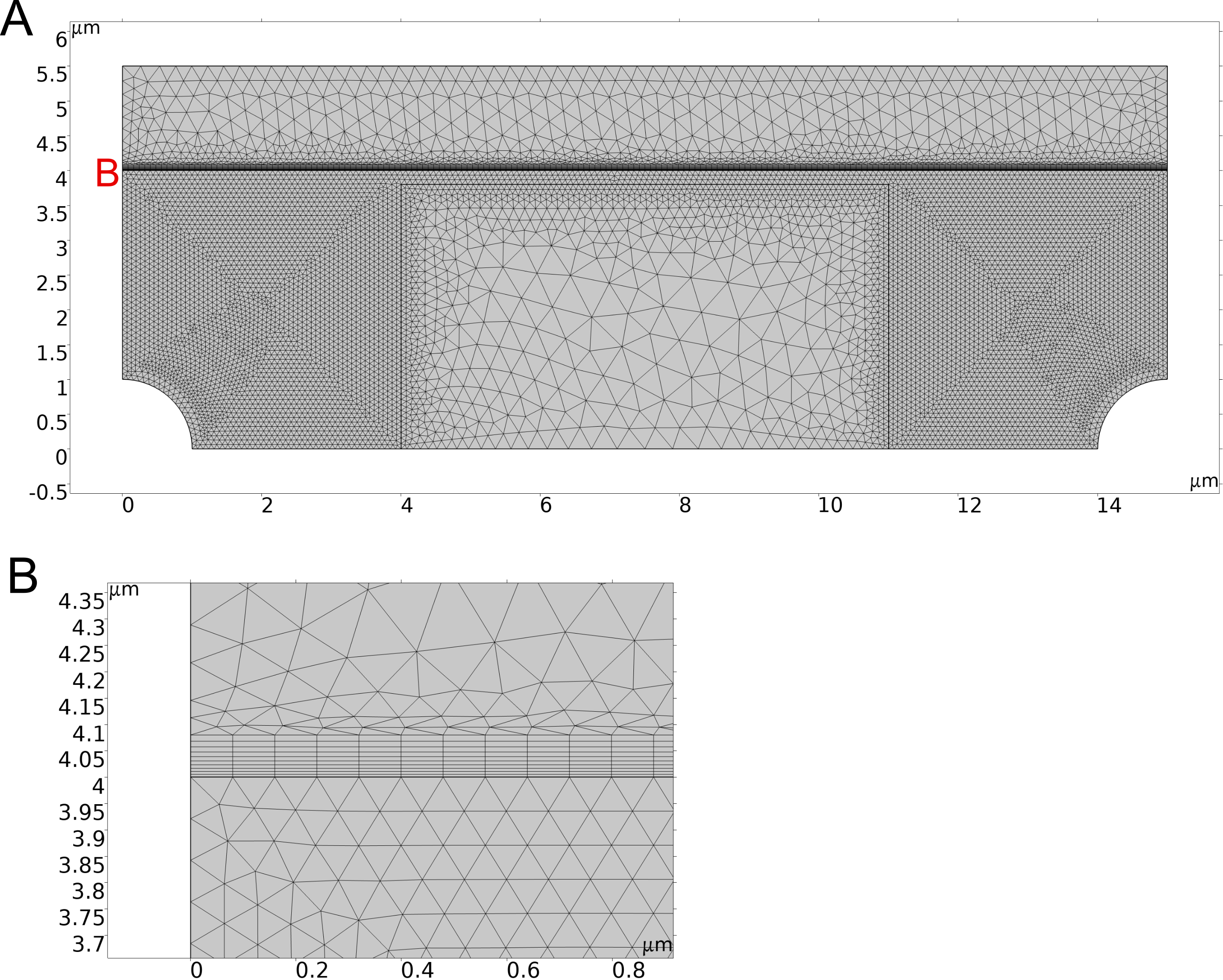}
\caption{\label{figureS1} \textbf{Non-uniform mesh used to simulate the model electron-bifurcating junction.} (A) The mesh elements vary in size by orders of magnitude over the simulation domain, especially to resolve boundary layers at the electrolyte-semiconductor boundary. (E) Three mesh sizes were used in the solid-state regions. The smallest was used to resolve the semiconductor near the electrolyte, the medium size was used to resolve the remainder of the semiconducting domain, and the largest was used to resolve the insulating domain. }
\end{figure}

\begin{figure}[H]\centering
\includegraphics[scale=0.75]{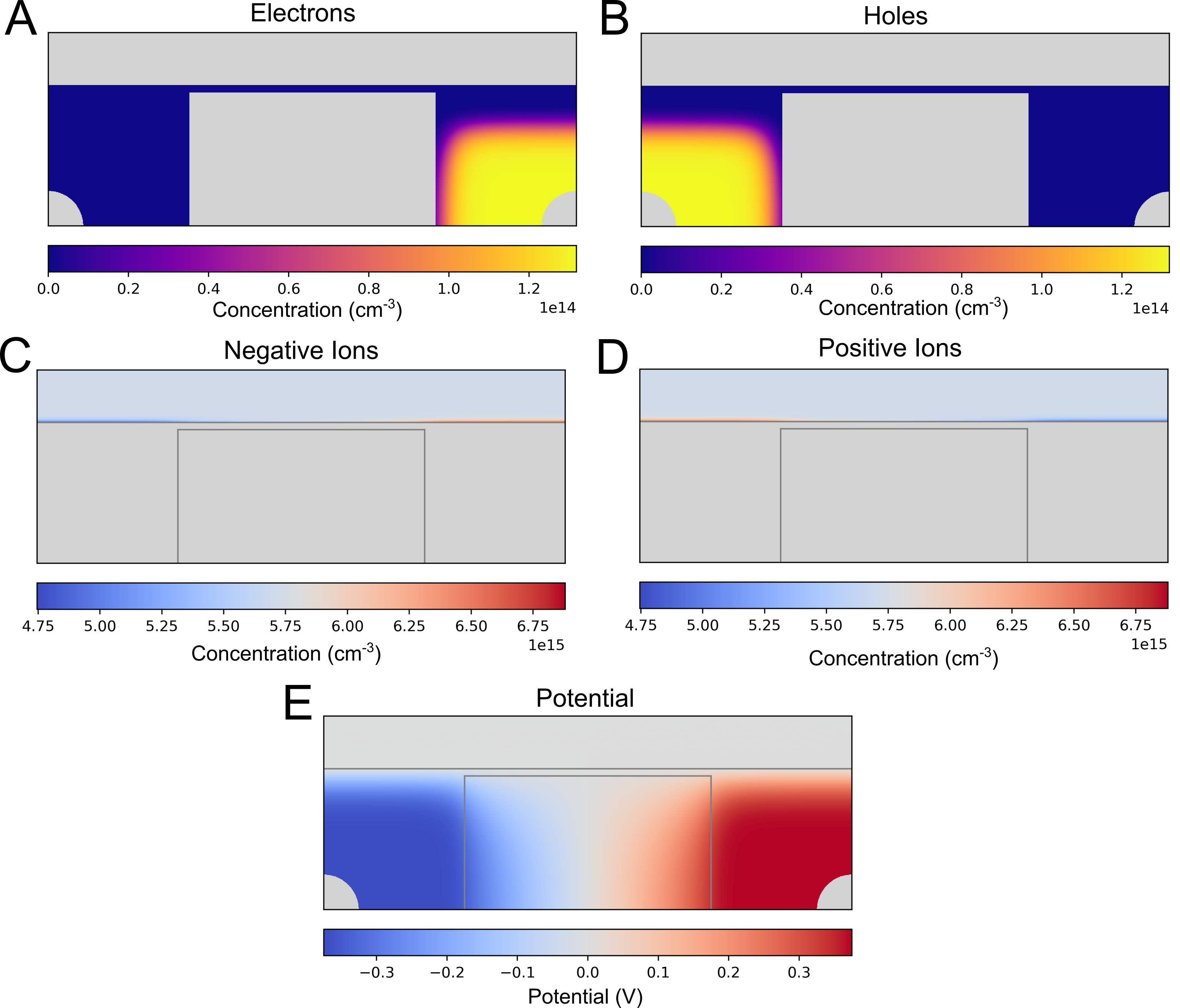}
\caption{\label{figureS2} Steady-state (A-B) Carrier distributions in the semiconductor, (C-D) ionic layers in the electrolyte solution, and (B) potential profile of the model bifurcating junction in Figure 3A for $V_{\text{bias}} = 0.1 $ V and $\mu_{s} = 3.4 \times 10^{-1} \text{ cm}^2\text{V}^{-1}\text{sec}^{-1}$.}
\end{figure}

Solving the above coupled equations over the specified domains is a highly nonlinear problem with boundary layers that must be resolved. Thus, a fine and highly nonuniform mesh is required\cite{selberherr2012analysis}. The mesh used is shown in Figure \ref{figureS1}A. Using the Boundary Layers mesh tool in COMSOL®\cite{comsol}, boundary layer meshes were introduced at the electrolyte-semiconductor interface domain to capture the thin Debye layer with mesh elements smaller than the Debye length (Equation \ref{Eq:Debye} and Figure \ref{figureS1}B-C). There are 10 boundary mesh elements, each larger than the previous by a factor of 1.1, with the first boundary elements having a width of $L_D/10$. The solutions for the ion concentrations forming these layers are shown in Figure \ref{figureS2}.

The meshes in the bulk of the semiconductor, insulator, and electrolyte were created using the Free Triangular mesh tool in COMSOL®. For the semiconductor, elements calibrated for Semiconductor were used with a max element size of $0.08$ $\mu$m, a min element size of $0.02$ $\mu$m, a maximum element growth rate of 1.1, a curvature factor of 0.25, and a narrow regions resolution of 1. For the insulator, elements calibrated for General Physics were used with a max element size of $0.6$ $\mu$m, a min element size of $0.2$ $\mu$m, a maximum element growth rate of 1.3, a curvature factor of 0.3, and a narrow regions resolution of 1. For the electrolyte bulk (past the boundary elements described above), elements calibrated for General Physics were used with a max element size of $0.5$ $\mu$m, a min element size of $0.05$ $\mu$m, a maximum element growth rate of 1.3, a curvature factor of 0.3, and a narrow regions resolution of 1. 

To solve the coupled multiphysics problem, several COMSOL® Multiphysics interfaces were used: Semiconductor, Transport of Charge Carriers (for the electrolyte), and Electrostatics. Within the Semiconductor interface, the log formulation was used with linear finite-element shape functions, and for the Transport of Charged Carriers module, I used the log formulation with quadratic finite-element shape functions (for robust convergence). For the Electrostatics interface, quadratic finite-element shape functions were used.

\section{Performance and carrier mobility}

Since the right and left metal contacts are assumed to be perfectly selective, and every initial charge injected by $\ce{DH-}$ is assumed to be followed up by a charge from $\ce{D^\cdot -}$ to the conduction band (Equations \ref{Eq:pelectrolyteboundary} and \ref{Eq:nelectrolyteboundary}), then all currents into the junction can be calculated from two quantities. First, the integrated charge injected from the electrolyte
\begin{equation}
    I_{\text{inj}}/A = \frac{1}{A} \int_S dS \hspace{3 pt }2|e^-|R_{\text{init}} = \int_S dS \hspace{3 pt}  p \hspace{2 pt} c_{\ce{DH^-}} \hspace{2 pt} k_{\ce{DH^-}\rightarrow v} 
\end{equation}
where the integral is over the surface $S$ of the electrolyte-semiconductor interface with area A. The division by $A$ normalizes the quantity for comparison with the total recombination rate
\begin{equation}
    I_{\text{rec}}/A = \frac{1}{A}\int_V dV \hspace{3 pt} R_{\text{rec}} = \frac{2|e^-|}{A}\int_{V} dV \hspace{3 pt} p \hspace{1 pt} n \hspace{2 pt} k_{\text{Langevin}},
 \end{equation}
where $V$ is the volume of the semiconductor. The quantities above were calculated using the COMSOL® Surface Integration and Line Integration functions, and are plotted in Figure 3B of the main text. Given the assumptions above, one can calculate the current into the left (L, hole-selective) and right (R, electron-selective) contacts, $I_{L}$ and $I_{R}$, respectively:
\begin{equation}
    I_{L} = \frac{I_{\text{inj}}}{2} - T_{\text{rec}}, 
\end{equation}
\begin{equation}
    I_{R} = \frac{I_{\text{inj}}}{2} -T_{\text{rec}}.
\end{equation}

\begin{figure}[H]\centering
\includegraphics[scale=0.45]{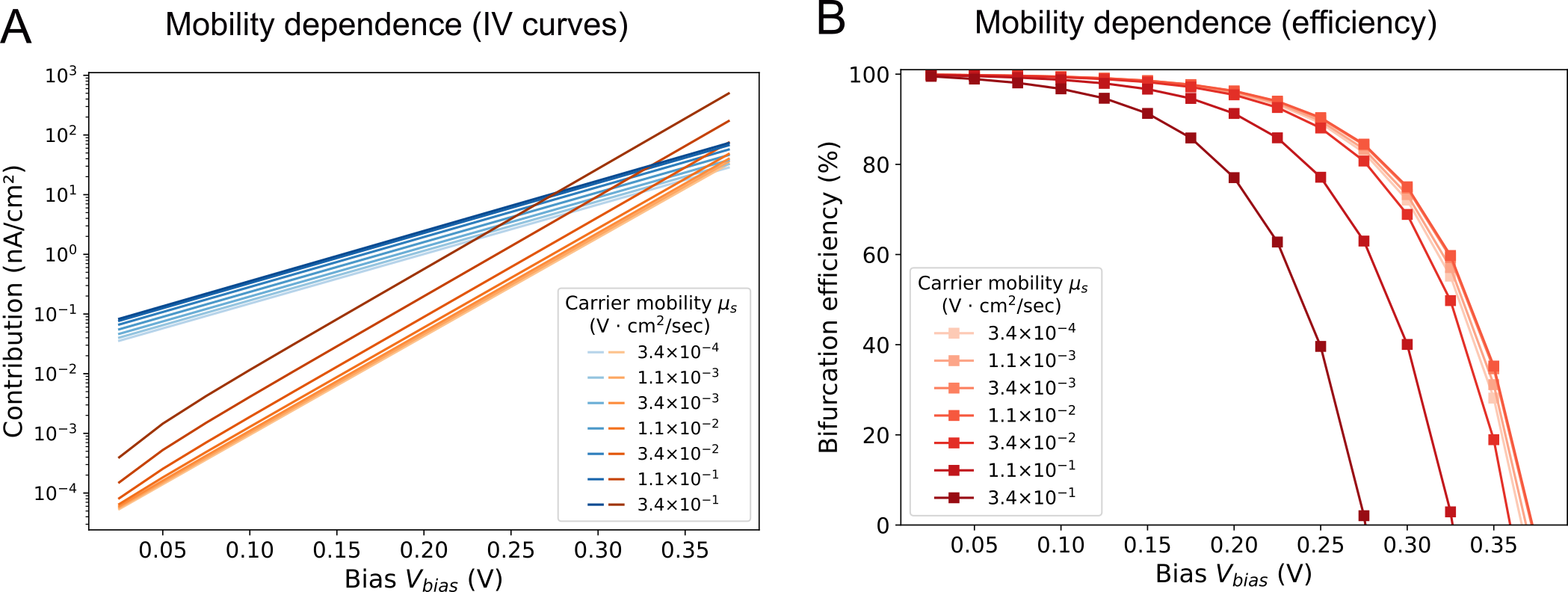}
\caption{\label{figureS3} \textbf{Simulated junction performance over a range of carrier mobilities $\mu_s$}. (a) Calculations of steady state currents $I_{\text{inj}}/A$ (oranges) and $I_{\text{rec}}/A$ (blues). These IV curves are approximately constant until $\mu_s > \mu^{\text{crit}}_s$ where $\mu^{\text{crit}}_s \approx 10^{-1} \text{ cm}^2\text{V}^{-1}\text{sec}^{-1}$. At higher carrier mobilities (B) the bifurcating efficiency (reds) of the junction at high bias $V_{\text{bias}}$ decreases significantly. }
\end{figure}

Steady state calculations were performed across a range of carrier mobilities $\mu_s$, and the results are shown in Figure \ref{figureS3}. The performance of the junction is optimized when $\mu_s< \mu^{\text{crit}}_{s}$ where $\mu^{\text{crit}}_s \approx 10^{-1} \text{ cm}^2\text{V}^{-1}\text{sec}^{-1}$. However, efficiency remains roughly constant for mobilities lower than $\mu^{\text{crit}}_{s}$. Thus, for the junction design in Figure 3A, low mobility materials ($<\mu^{\text{crit}}_{s}$) should be used, but no further fine-tuning of the mobility is useful.

\bibliographystyle{pnas.bst}
\bibliography{bibliography}